\documentclass{PoS}

\title{The scalar does not decay at finite temperatures}

\ShortTitle{Screening masses of mesons}

\author{\speaker{Debasish Banerjee}, Rajiv Gavai and Sourendu Gupta\\
        Department of Theoretical Physics, Tata Institute of Fundamental Research,\\
        Homi Bhabha Road, Mumbai 400005, India\\
        E-mail: \email{debasish@theory.tifr.res.in,gavai@tifr.res.in,sgupta@tifr.res.in}}

\abstract{We investigate medium effects on mesonic screening lengths for QCD with
          2-flavours of dynamical staggered quarks on lattices with cutoff a=1/6T.
          Denoting the cross-over temperature by $T_c$, we vary the temperature T         
          from 0.89$T_c$ to 1.92$T_c$, spanning both the hadronic and quark gluon
          plasma phases. While chiral symmetry restoration in the vector channel
          appears to take place near $T_c$, it is seen in the scalar channel only
          above 1.33 $T_c$. Varying spatial lattice sizes, we find very little volume
          dependence in our results at 0.94 $T_c$. We discuss the stability of the
          scalar meson at these temperatures.}

\FullConference{The XXVIII International Symposium on Lattice Field Theory, Lattice2010\\
		June 14-19, 2010\\
		Villasimius, Italy}

\begin{document}

\section{Introduction}
When strongly interacting matter is heated, it is expected to go over to the Quark
Gluon plasma (QGP) phase. Lattice Quantum chromodynamics (QCD) is extensively used
to study the properties of matter across this transition. Even though the thermodynamic
quantities
and the equation of state is known in sufficient detail, more information about the 
plasma phase is needed to distinguish it from the hadronic phase.

The study of static correlation lengths constitute one such important clue. The low lying
degrees of excitations of the plasma can be studied using the spatial correlation
function of hadronic operators \cite{DeTarKogut}. The screening mass is obtained from the
exponential decay of the screening correlators at large spatial separations. Under certain
analyticity assumptions, this screening mass can be related to the 
pole of the real time propagator of the corresponding operator as argued in \cite{DeTarKogut}.
These correlation functions, and the screening masses encode
information about the interactions in the medium, indicating for example, whether
certain symmetries broken at low temperatures are restored at high temperatures.
It is known that at very high temperatures the
screening masses of the mesons approach that of a free ideal gas, 2$\pi$T \cite{bornetal}.
Moreover, finite volume corrections for the thermodynamic quantities can also be
studied using the screening masses. 

We investigate the spatial correlation functions for hadrons in 2-flavour QCD with 
staggered fermions in both the hadronic and the plasma phases. We study the pattern
of chiral symmetry restoration across the cross-over in the screening
spectrum. We study the finite volume effects on the correlation functions at
low temperature and the possibility of the decay of the scalar meson.

\section{Technical Details}
The generation of the configurations used in the present study was reported in
\cite{rvgsgNt6}. The lattices had a temporal extent of $N_t=6$ and spatial extent
$N_s=24$, except in the finite volume study where spatial lattice sizes from 8 to
30 were used. Two flavours of light staggered quarks were used with the bare quark
masses tuned
such that the $T=0$, $m_{\pi} \approx 230$ MeV at all temperatures T.
\newline
Point-to-point correlation functions for local meson operators
in the pseudo-scalar (PS), scalar (S), vector (V) and
axial-vector (AV) channels were analyzed in detail.
Denoting by $G(x,0)$ the staggered propagator, 
the correlation functions projected to zero momentum are:
\begin{equation}
 C(z) = \frac{1}{N_x^2 N_t} \sum_{x,y,t} \textmd{Tr} [G(x,0) G^{\dagger}(x,0)] g_S(x)
\end{equation}
where $g_S(x)$ \cite{rvgsgpm}
are the staggered phase factors specific to the mesons in 
given quantum number channels. We have calculated only
the connected part of the correlation functions, which correspond to
the flavour non-singlet mesons. These correlation functions
contain contributions from the respective parity 
partner as well, and hence are parameterized as
\begin{equation}
\label{2masfit}
 \nonumber C(z) = A_1(~\mathrm{e}^{-m_1 z} + \mathrm{e}^{-m_1(N_z-z)}~) 
      + (-1)^z A_2(~\mathrm{e}^{-m_2 z} + \mathrm{e}^{-m_2(N_z-z)}~) 
\end{equation}
Here $m_1$ and $m_2$ are the screening masses of the lightest
natural parity meson appropriate for the operator used and its opposite 
parity partner.  $A_1$ and $A_2$ are the respective normalizations. 
In this convention, the Goldstone pion is the non-oscillating 
part of the correlator and has positive $A_1$. 


 The tolerance of the conjugate gradient (CG) algorithm was chosen to be  
$\epsilon = 10^{-5}$. This ensured that the systematic error in the
meson correlation functions arising due to the matrix inversion
was much lesser than the statistical errors.

The screening masses were extracted by fitting the correlation functions to the 
2-mass fit form in eqn \ref{2masfit}. As usual, the correlation between the different
z-slices are incorporated through the covariance matrix and the fit values
are obtained by minimizing the correlated-$\chi^2$.

\section{Results}
While our main results are on the temperature dependence of the screening
mass spectrum on $N_t=6$ lattices, and the consequent study of the chiral
symmetry restoration pattern as a function of temperature, we also present
results for the spatial volume (in-)dependence of the screening masses of
all the mesons near the critical end-point temperature.

\begin{figure}[!tbh]
\begin{center}
\hbox{
\includegraphics[width=0.5\textwidth]{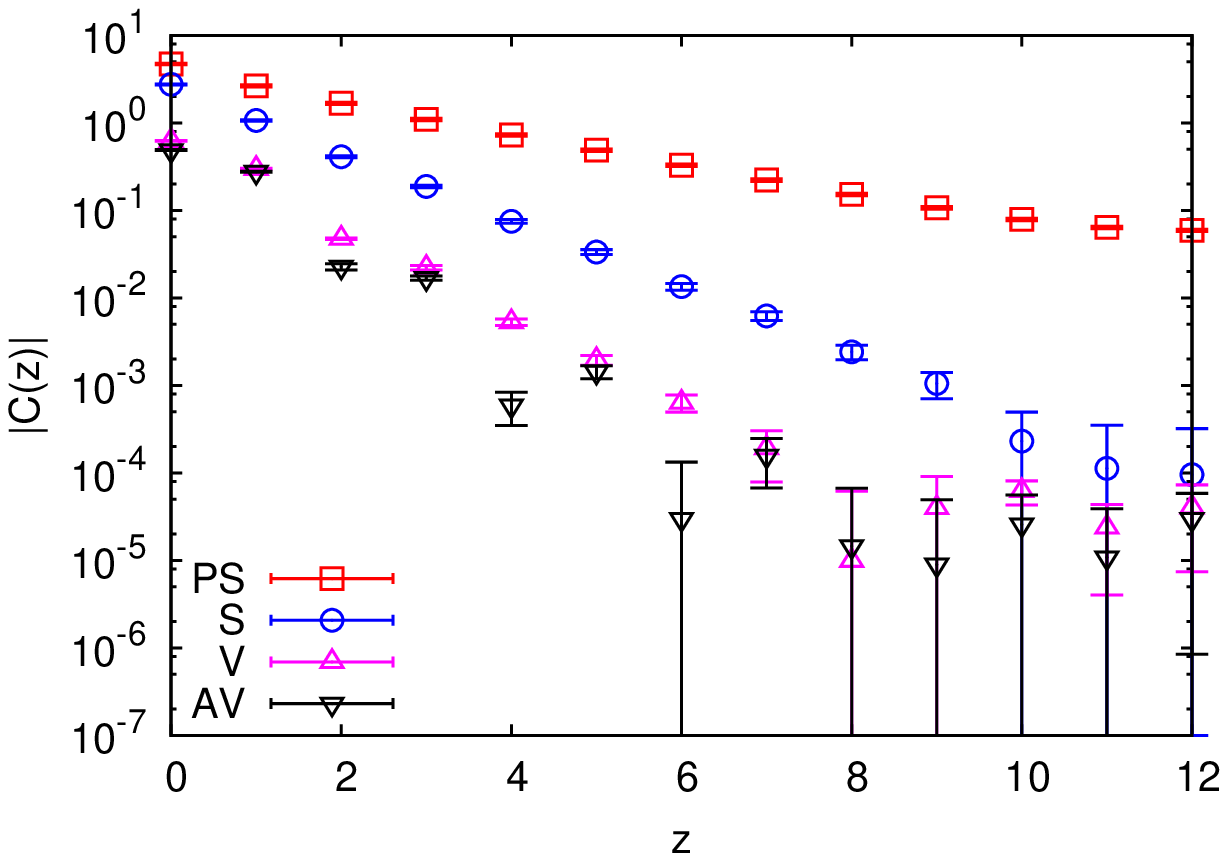}
\includegraphics[width=0.5\textwidth]{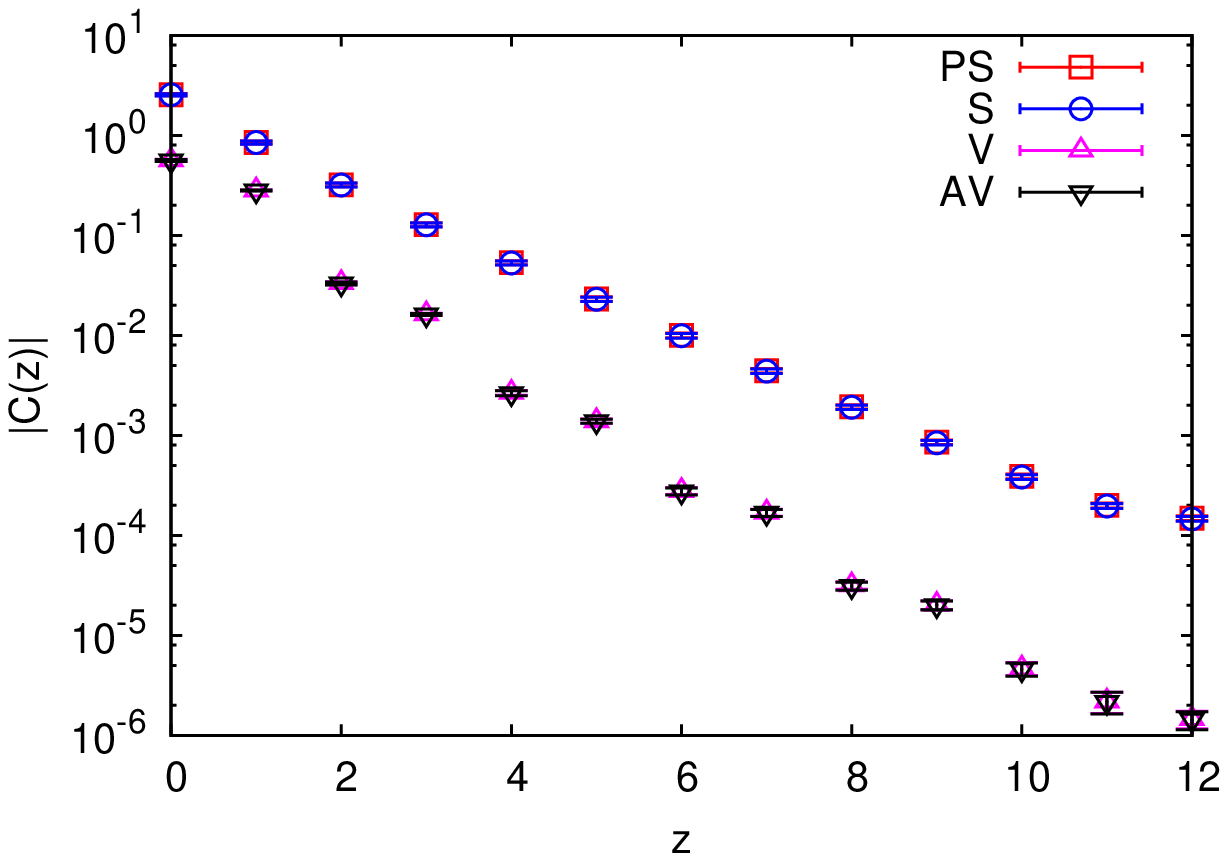}
}
\end{center}
\caption{The correlators for all the mesons
at T=1.92 $T_c$(left) and T=0.97 $T_c$(right).
Note the prominent oscillations for the V-AV
mesons. We have plotted the absolute 
value of the correlators for the V and AV mesons.}
\label{fig:corrl}
\end{figure}

\subsection{Screening spectrum}
We display our results for all the meson correlators in fig \ref{fig:corrl} at two 
different temperatures: 1.92 $T_c$ and 0.97 $T_c$. The V/AV correlators 
show prominent oscillations at both temperatures. For the same statistics, there is
more noise for the other mesons at large-z and lower temperature than the pion, which
is equally good at both temperatures. The pion correlation functions could be 
fit well to a single mass form at all temperatures, except the highest temperature 
(1.92 $T_c$) where we had to use the 2-mass fit. A full list of the 
screening masses will be given elsewhere\cite{dbrvgsg}.

\begin{figure}[!tbh]
\begin{center}
\hbox{
\includegraphics[width=0.5\textwidth]{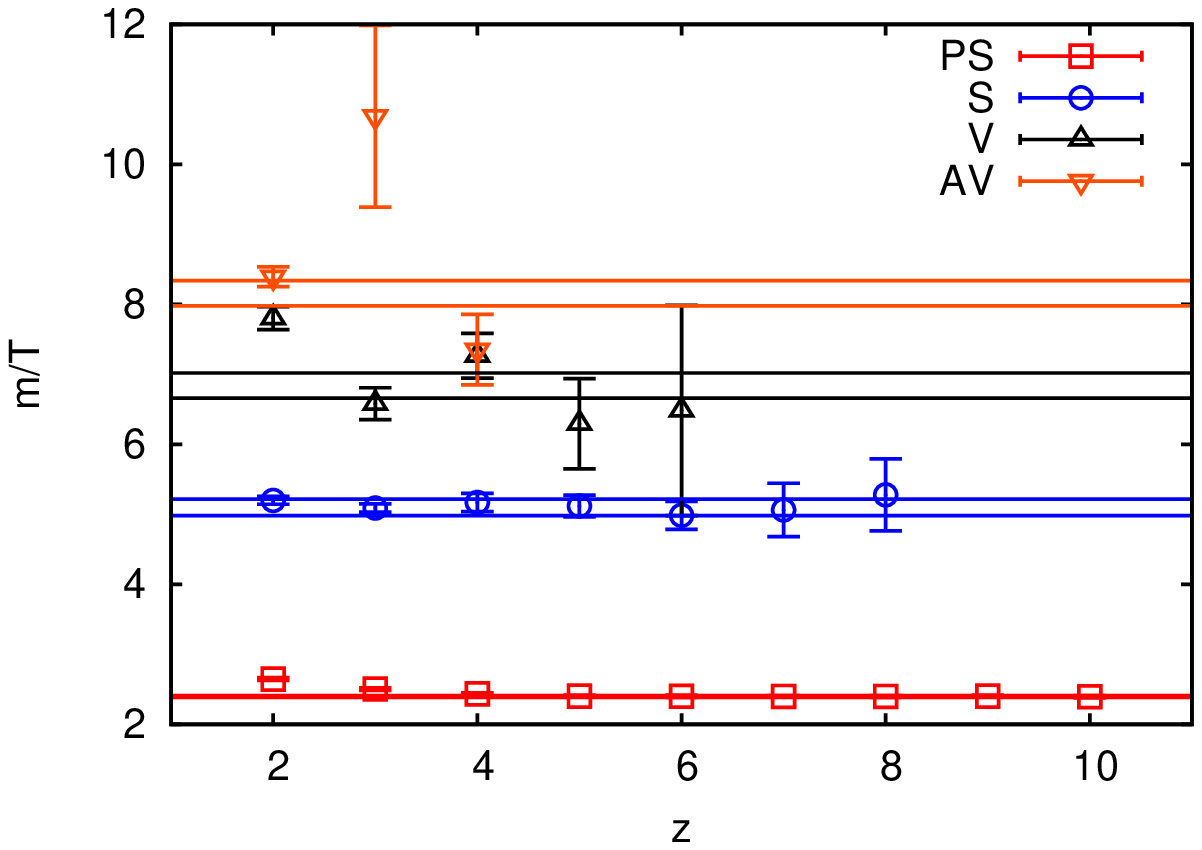}
\includegraphics[width=0.5\textwidth]{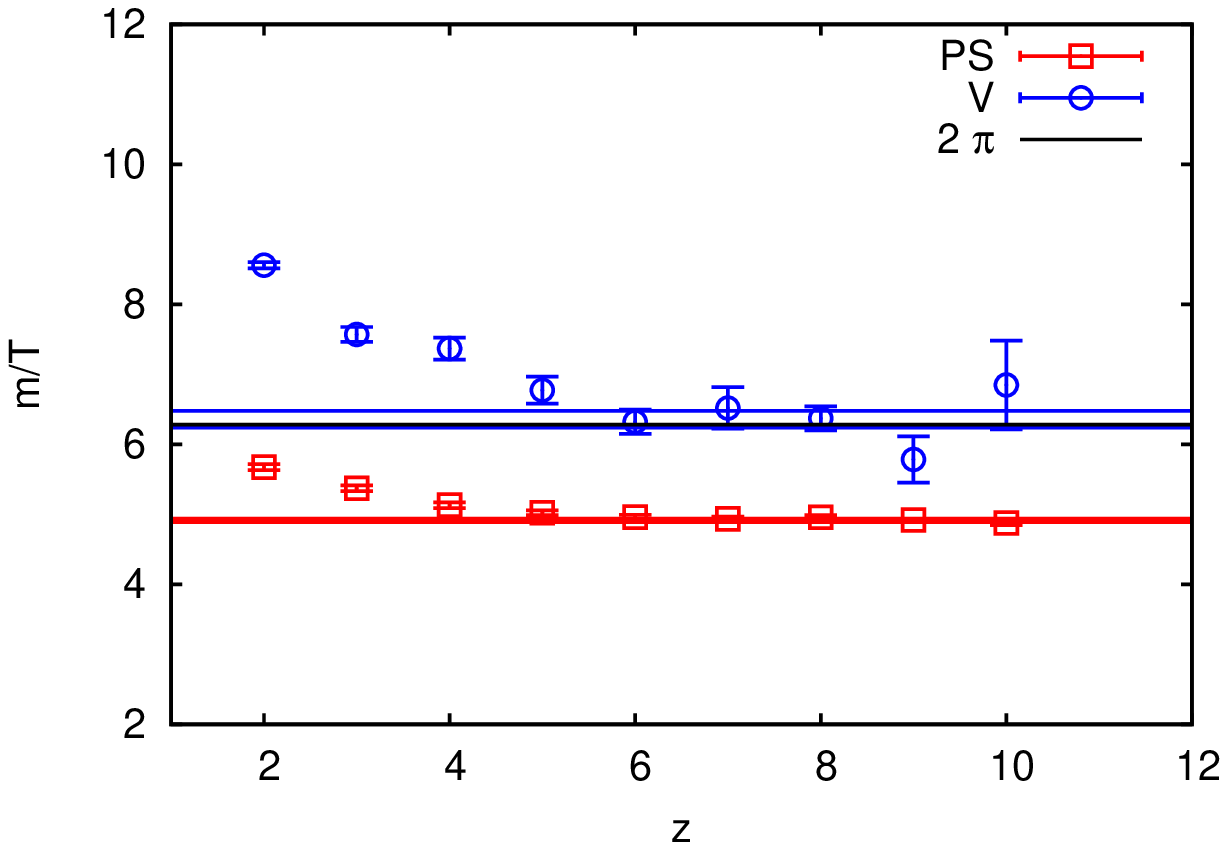}
}
\end{center}
\caption{The extracted local masses for the mesons at
T=0.97 $T_c$(left) and T=1.92 $T_c$(right). For T=1.92 $T_c$,
only the PS and V meson masses are shown since the S and AV
exactly coincide with them. The black line coinciding with the
lower blue line is the ideal gas limit.
}
\label{fig:localm}
\end{figure}

To check the consistency of fitted masses, we also calculated the local masses. Due to the
oscillating nature of the staggered fermions, the local masses were defined using
correlators separated by 2 z-slices as done in \cite{rvgsgpm}:
\begin{equation}
 \frac{C(z+1)}{C(z-1)} =
 \frac{\cosh[-m(z)(z + 1 - N_z/2)]} {\cosh[-m(z)(z - 1  - N_z/2)]}~.~
\end{equation}
The local masses so extracted are shown in 
Fig. \ref{fig:localm} for the same two temperatures, 1.92 $T_c$ and 0.97
$T_c$. The pion masses are the best determined in both the phases with small
error bars and agree very well with the fitted values. Such an agreement are
also found for the other mesons, although with larger bars.

\begin{figure}[!tbh]
\begin{center}
\hbox{
\includegraphics[width=0.5\textwidth]{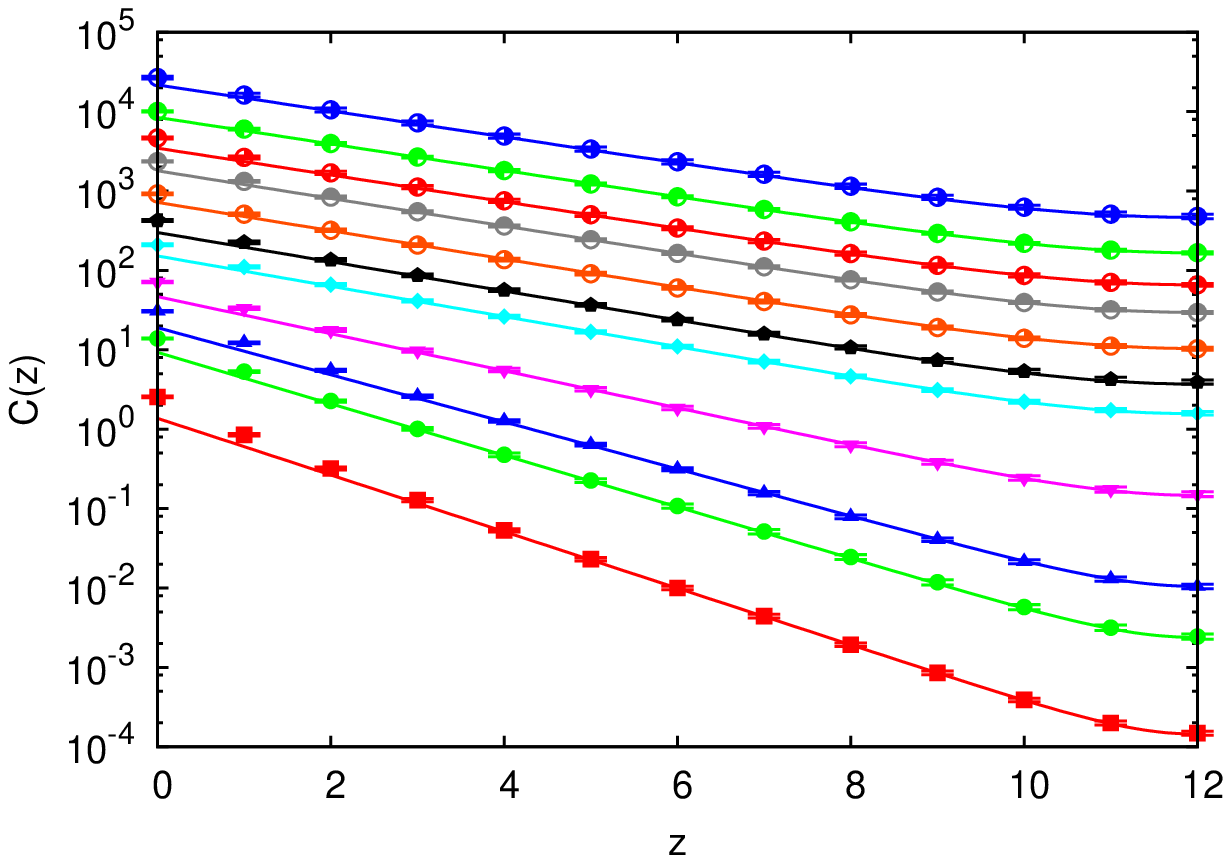}
\includegraphics[width=0.5\textwidth]{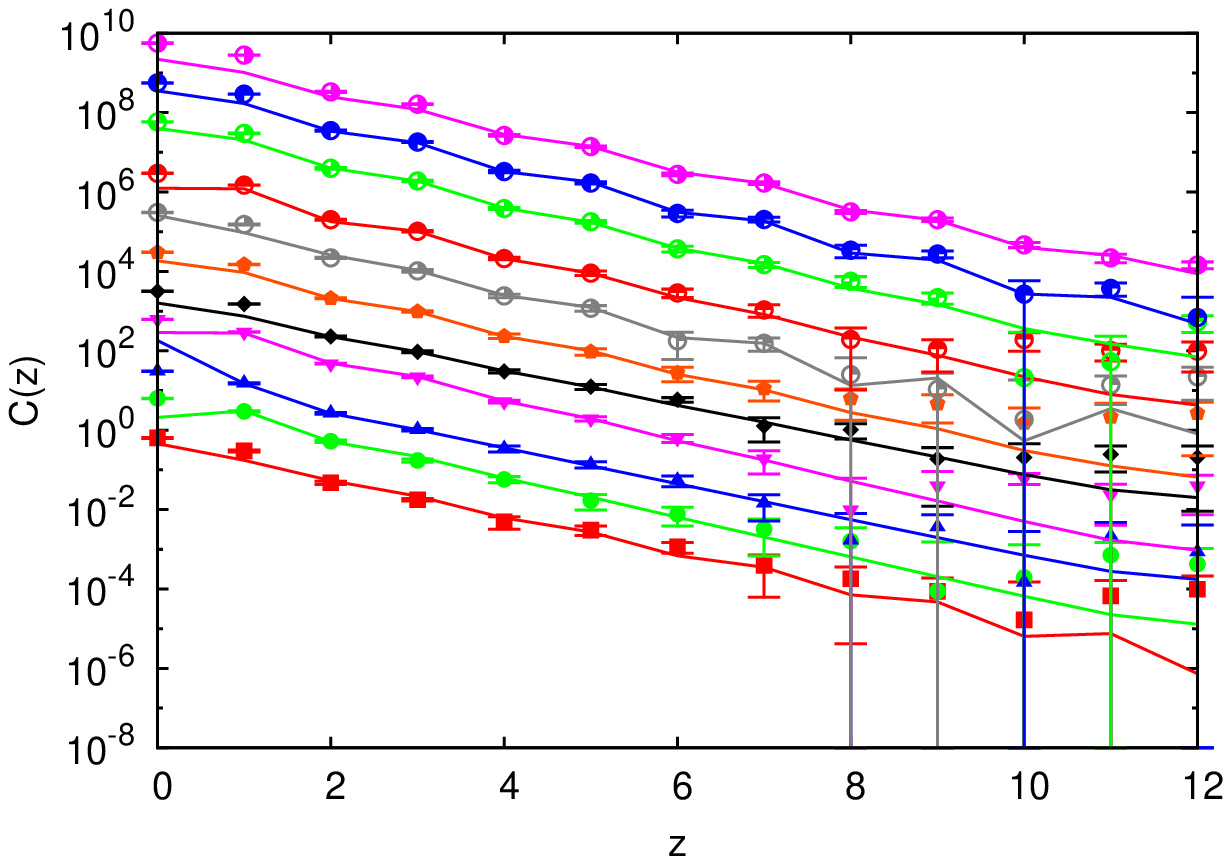}
}
\end{center}
\caption{Single mass fits to the pseudo-scalar correlators (left)
and the 2-mass fits to the vector correlators (right) for the entire temperature range
studied here. Correlators have been appropriately scaled to make
them fit in the same figure. The lines and symbols in each plot
represent temperatures in units of $T_c$ in decreasing order
from bottom: 1.92, 1.48, 1.33, 1.21, 1.01, 1.00, 0.99, 0.97, 0.94, 0.92,
0.89. (The lowest curve is for 1.92 $T_c$ and the topmost is 
for 0.89 $T_c$.)}
\label{fig:PSfits}
\end{figure}

The behaviour we discussed so far for both the hadronic and QGP phase
is generic for other temperatures as well, with gradual changes in the
fit parameters. Fig. \ref{fig:PSfits} shows the fits to the PS and V
correlators using the fitted values of the screening masses and corresponding 
amplitudes. The correlators have been scaled to depict the entire range in 
temperature on the same plot. The 2-mass nature of the fit used for the V 
correlators is very evident.

\begin{figure}[!tbh]
\begin{center}
\includegraphics[width=0.6\textwidth]{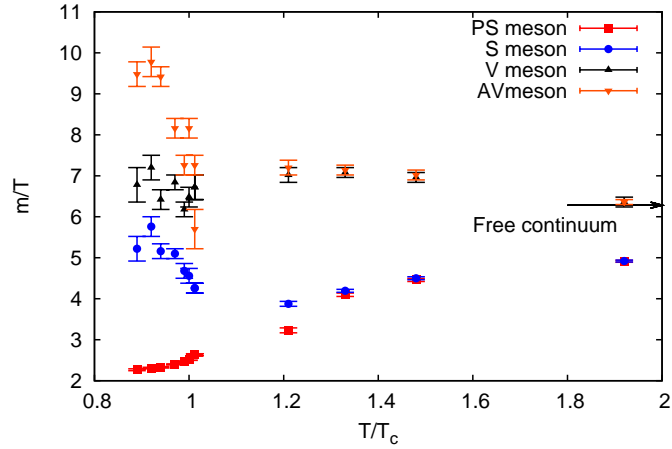}
\end{center}
\caption{Masses of the PS, S, V and AV mesons a a function of $T/T_c$.
The free continuum value of 2$\pi$ is indicated with an arrow.}
\label{fig:allmass}
\end{figure}

Finally, we plot the screening masses of the S, PS, V and AV mesons as a function
of $T/T_c$ in Fig. \ref{fig:allmass}. The chiral symmetry, broken at
low temperatures, gets restored sufficiently away from the cross-over temperature
$T_c$. While the PS and S mesons masses become degenerate only around 1.33 $T_c$,
the V and AV mesons are degenerate even at temperatures $\simeq T_c$. 
We note from Fig.\ref{fig:localm} that while the V/AV screening masses appear to be the same as the
free theory continuum limit at about 2 $T_c$, the screening masses of the
PS/S mesons still differ from the ideal gas, and the V/AV 
mesons, by about $\sim$ 20\%.
Our results confirm similar results obtained by other collaborations\cite{bornetal,rvgsgpm,swagato}.

\begin{figure}[!tbh]
\begin{center}
\includegraphics[width=0.6\textwidth]{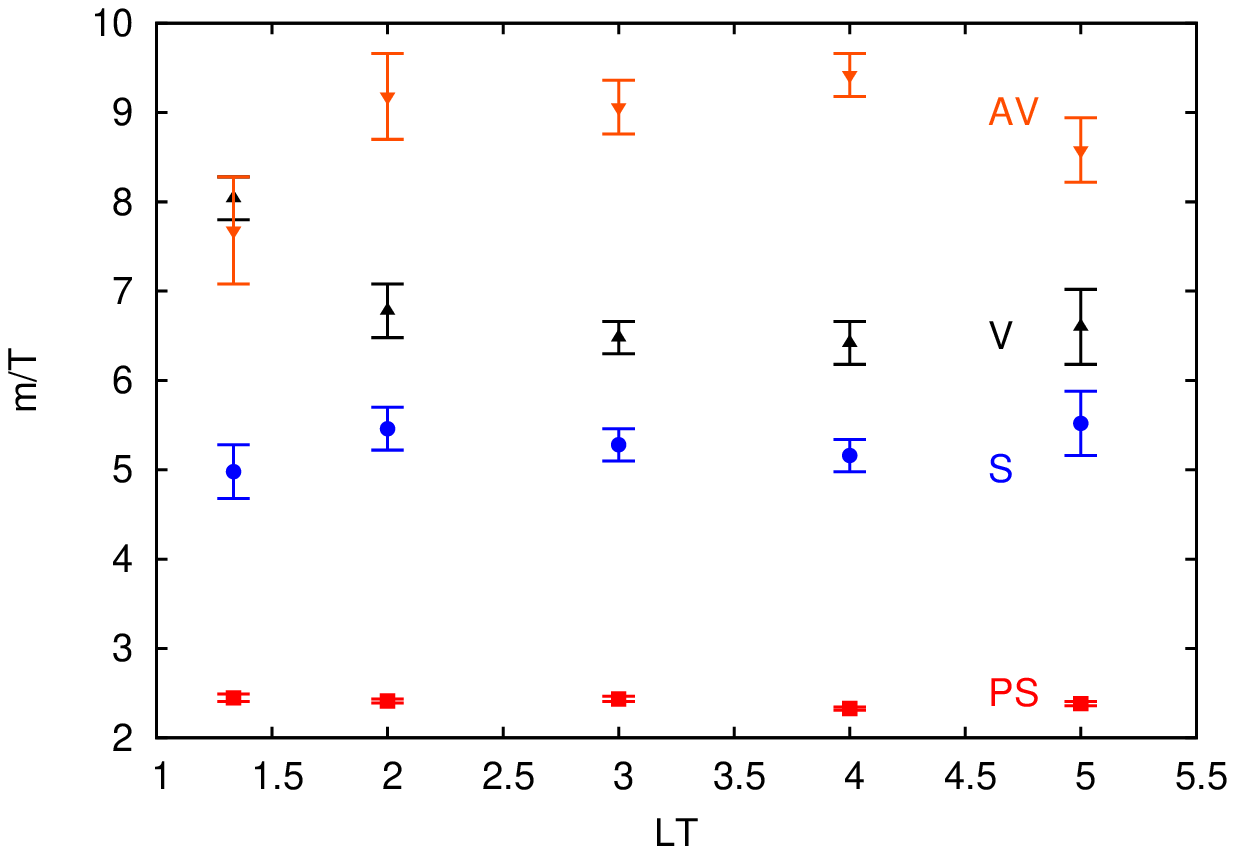}
\end{center}
\caption{Masses of the PS, S, V and AV mesons a a function of $LT$ at $T=0.94 T_c$.}
\label{fig:masV}
\end{figure}

\subsection{Finite Volume Study}
A finite volume study was done at 0.94 $T_c$, since this is the phenomenologically 
interesting temperature where the QCD CEP is expected to lie at $\mu_E=1.6 T_c$ \cite{rvgsgNt6}.
A range of lattices with $N_s$=8,12,18,24 and 30 were considered, where
the smallest two had anisotropic spatial extents. From fig. \ref{fig:masV}, it is clear
that the screening masses do not have significant volume dependence. All the lattices
in our study are much bigger than the corresponding screening lengths of the mesons,
which explains the absence of the finite volume effect associated with stable states.

\begin{figure}[!tbh]
\begin{center}
\hbox{
\includegraphics[scale=0.6]{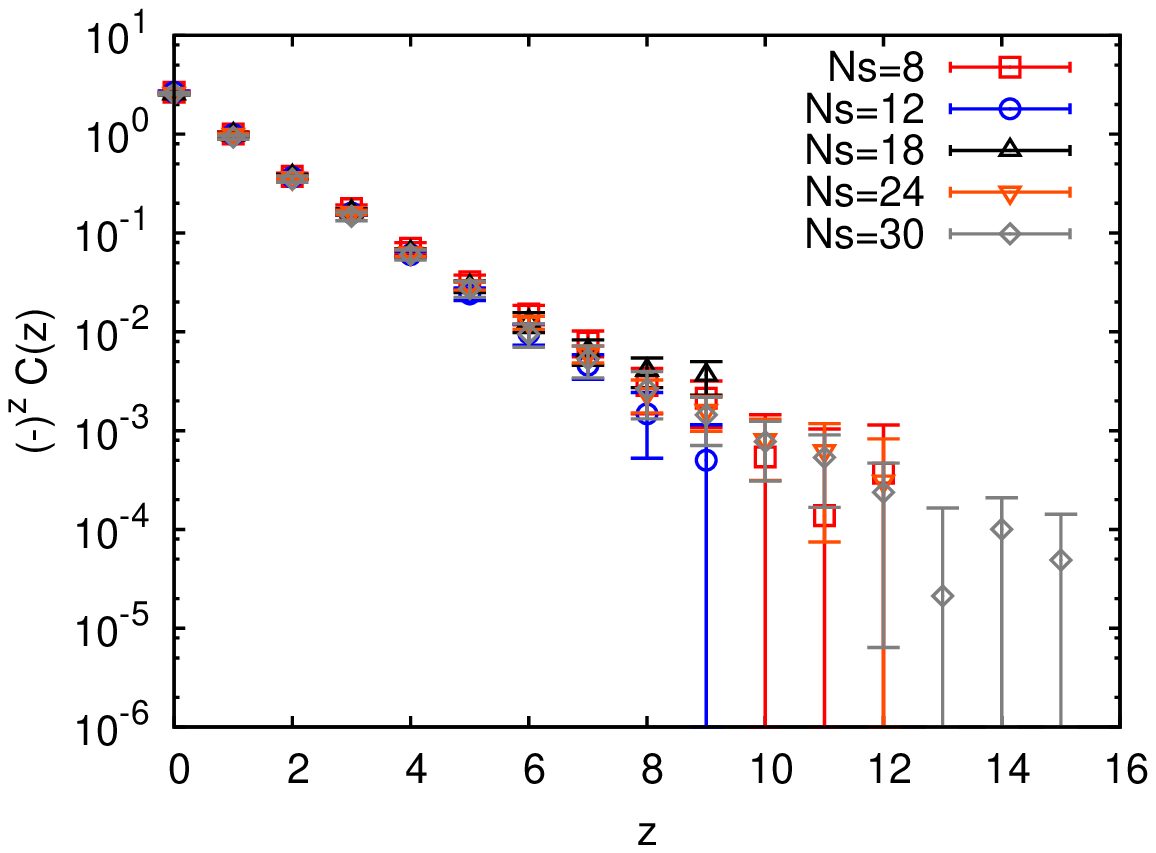}
\includegraphics[scale=0.6]{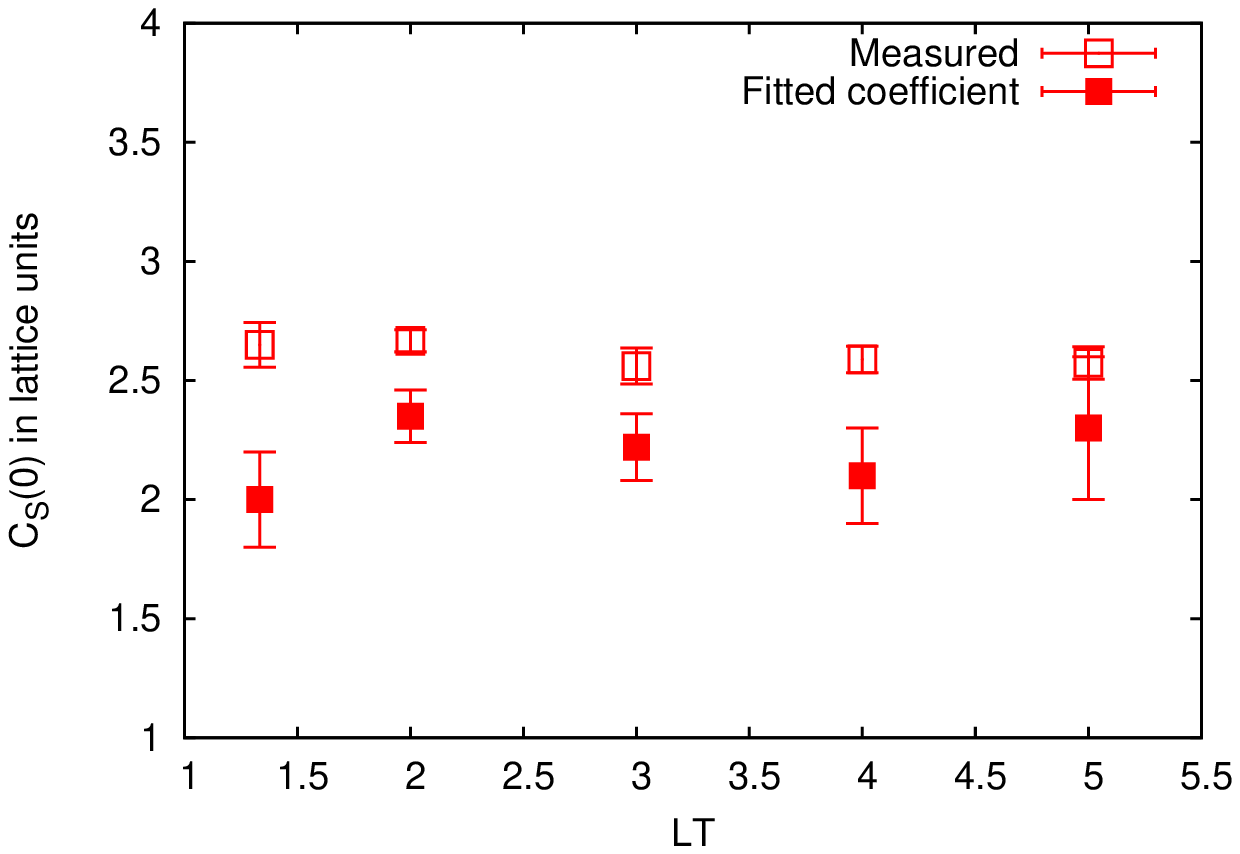}
}
\end{center}
\caption{The upper figure shows the Scalar meson correlation 
functions. It is quite featureless as a function of the volume. 
The lower figure shows that both the measured and fitted value of 
$C_S(0)$ is independent of volume.}
\label{fig:decay}
\end{figure}

Further, given that the mass of the scalar meson is more than twice the mass of the pion,
we can discuss whether the decay of the scalar is possible. In the continuum, this decay is
forbidden by parity. On the lattice however, unphysical decays are known to occur \cite{clb}.
In our case, parity and kinematical considerations imply that the scalar could only decay into
the pion and one of its taste partners.  A finite volume study could potentially answer this
interesting question. Our results indicate that this decay does not occur.
Fig \ref{fig:decay} shows the scalar correlation functions are quite featureless
as a function of volume. Further support comes from the fitted and the measured correlator
normalization, $C_S$(0), which are both independent of volume.

It is known that decays can be
arrested if the volume is too small. In this context, we
point out that we have worked with reasonably large volumes.
The LT values in our study range from 1.33 to 5 while m$_{PS}$L range from
2.4 to 12.


\section{Summary}
We studied point-to-point correlation functions for mesons in 
the PS, S, V and the AV channels for two flavour
QCD in the staggered fermion formulation. We extracted the corresponding
screening masses and showed that while the screening masses of the
V/AV mesons become degenerate by $T_c$, one needs to go up to 1.33 $T_c$ to see the corresponding 
degeneracy of the PS/S mesons. 
Moreover, while by 2 $T_c$ the V/AV screening masses 
are already seen to reach their continuum value,
the PS/S screening masses are still 15-20\% away
from the free theory value. Using a finite volume study, we find that the scalar
correlation functions are insensitive to the change in spatial volume, which leads us
to suspect that the scalar meson decay does not occur at
finite lattice spacings at temperatures less than the cross-over temperature.

\section*{Acknowledgments}
The computations were performed on the CRAY X1 of the
Indian Lattice Gauge Theory Initiative (ILGTI) in TIFR, Mumbai. D.B.
wishes to acknowledge useful discussions with Saumen Datta, 
Nilmani Mathur and Jyotirmoy Maiti.


\begin{thebibliography}{99}
\bibitem{DeTarKogut}
C. \ DeTar and J. B. \ Kogut {\sl Phys.Rev.D36(1987) 9}
\bibitem{bornetal}
K. Born et al (MTc Collaboration) {\sl Phys. Rev. Lett. 67, 302 (1991) }
\bibitem{rvgsgNt6}
R. V. \ Gavai, S. \ Gupta {\sl Phys.Rev.D78 (2008) 114503}
\bibitem{rvgsgpm}
R. V. \ Gavai, S. \ Gupta and P. \ Majumdar {\sl Phys.Rev. D65 (2002) 054506}
\bibitem{dbrvgsg}
D. \ Banerjee, R. V. \ Gavai and S. \ Gupta {\sl (in preparation)}
\bibitem{swagato}
S. \ Mukherjee {\sl PoS LAT2007:210,2007}
\bibitem{clb}
C. \ Bernard et al {\sl Phys.Rev. D76 (2007) 094504 }
\end{thebibliography}
\end{document}